# Revelation of Double Magicity in N=Z Nuclei in the RP-Process Region

M.M. Sharma[a] and J.K. Sharma[b]

[a]Physics Department, Kuwait University, Kuwait 13060, Kuwait
[b]Physics Department, St. John's College, Agra-282002, India

**Abstract.** In rapid-proton capture (*rp*-process), N=Z nuclei above Ni are understood to act as waiting-point nuclei. The N=Z nuclei $^{68}$Se, $^{72}$Kr, $^{76}$Sr and $^{80}$Zr among others are known to give rise to a large-energy x-ray flux and peaks in abundances of these nuclei synthesized in the astrophysical *rp*-process. Investigating the experimental isotope shifts in Kr isotopes near the proton drip-line within the framework of the deformed Relativistic Hartree-Bogoliubov theory, we have discovered that N=Z *rp*-process nuclei $^{68}$Se, $^{72}$Kr, $^{76}$Sr and $^{80}$Zr exhibit large shell gap both at the proton and neutron numbers in the deformed space with the consequence that pairing correlations for protons and neutrons vanish. This lends a doubly magic character to these nuclei. A significant number of nuclei in this region are also shown to exhibit neutron magicity at N=34, 36, 38, and 40 in the deformed space. A unique case of concomitance of the double magicity and the shape-coexistence is found for $^{68}$Se.

## 1. INTRODUCTION

In rapid-neutron capture (*r*-process nucleosynthesis) nuclei with neutron magicity act as *waiting-points* [1]. A major shell gap at the magic numbers reduces the neutron capture probability significantly. Subsequently, these nuclei wait and undergo β-decay before capturing a neutron for further processing [2]. On the other hand, in rapid-proton-capture (*rp*-process) in x-ray bursters, N=Z nuclei above Ni (Z=28) are understood to play an important role in producing a part of x-ray flux [3-5]. These nuclei are deformed and have no known shell closure. A number of these nuclei are considered to be waiting-point due to inability of these nuclei to capture a proton, thus hindering the *rp*-process. Consequently, peaks have been shown to arise in abundances of the N=Z nuclei $^{68}$Se, $^{72}$Kr, $^{76}$Sr and $^{80}$Zr among others in the astrophysical *rp*-process in x-ray bursters [4].

Precise measurements [6-10] of nuclear masses have established the waiting-point nature of the N=Z nuclei $^{68}$Se, $^{72}$Kr, $^{76}$Sr, and $^{80}$Zr. Negative value of $Q_p$ for proton capture is cited as evidence for the waiting-point character of these nuclei. This is interesting, for these nuclei are highly deformed and no known magic number is present in this region in sharp contrast to the *r*-process waiting-point nuclei which arise due to the major magic numbers, viz. N= 50, 82 or 126. This poses a theoretical challenge to understand the structure of the waiting-point character of the *rp*-process nuclei responsible for x-ray bursts.

The structure of these nuclei is far from clear. What is known is that nuclei in this region are highly deformed [11-13] until the major magic number approaches at Z=N=50 where the *rp*-process is predicted to terminate [4]. In this talk, we present our study of the experimental isotope shifts in proton-rich Kr nuclei near the *rp*-process path within the framework of the relativistic Hartree-Bogoliubov (RHB) theory in deformed space. It is shown that the negative isotope shift of a nucleus such as $^{72}$Kr stems from vanishing pairing correlations due to the creation of large shell gaps at the respective neutron and proton numbers. Consequently, the nucleus $^{72}$Kr is identified as doubly magic in deformed space. We also report on the doubly-magic character discerned for a number of other waiting-point nuclei along the *rp*-process path, viz. $^{68}$Se, $^{76}$Sr, and $^{80}$Zr. We find that several nuclei in the *rp*-process region exhibit magicity of neutrons in the deformed space [14].

## 2. THE ISOTOPE SHIFTS IN KR NUCLEI

Isotope shifts in nuclei reveal shell effects across a major shell gap [15]. The most prominent case in this context is that of the Pb isotopes, where one observes a conspicuous kink at the magic number N=126 in the measured isotope shifts [15]. Efforts aimed at describing the observed kink within the approach of the nonrelativistic density-dependent Skyrme approach have not met success [16]. With the advent of the RMF theory for finite nuclei, it was shown for the first time that the RMF force NL-SH [17] reproduced the anomalous kink in the isotope shifts of Pb chain successfully [18]. A further investigation of the differences between the nonrelativistic Skyrme approach and the RMF theory has revealed that the isospin dependence of the spin-orbit potential which is produced as a result of the Dirac-Lorentz structure of nucleons in the RMF theory is responsible for the kink in the isotope shifts of Pb nuclei [19].

For other isotopic chains such as Kr and Sr, the shell effects in the isotopic shifts are shrouded by the deformation of nuclei. The experimental values (with N=50 as a reference point) for the Kr chain are known for isotopes down to N=36 with a significant precision from laser spectroscopy measurements [20] (see Fig. 1). The nuclide $^{72}$Kr (N=36) is very close to the proton drip line. The salient feature of these data is the monotonous increase of $\Delta r_c^2$ from N=50 down to N=40 due to an increasing deformation in going to the lighter isotopes. A similar feature is seen for isotopes heavier than N=50. However, the most striking feature that is demonstrated by the experimental data is that $\Delta r_c^2$ for $^{74}$Kr is nearly zero and that it becomes strongly negative for $^{72}$Kr. The magnitude of the isotope shift or in other words, the charge radius of a nucleus with respect to that of the reference point is mainly reflective of the deformation of a nucleus. To the first order, it is approximated as

$$\Delta r_c^2 = \langle r_c^2 \rangle \frac{5}{4\pi} \delta \beta_2^2 \,, \tag{1}$$

where $\Delta r_c^2$ is the mean-square charge radius of the spherical nucleus and $\beta_2$ is the quadrupole deformation of the deformed one. This excludes any effects due to shell structure. The downward trend of the experimental isotope shift below N=40 implies unquestionably that a structural factor is at play. In this work, we have investigated this factor that causes a decreasing or even a negative isotope shift.

## A. The Isotope Shifts in the RMF+BCS Theory

First, we have investigated the isotope shifts of Kr nuclei within the framework of the relativistic mean-field theory [21]. We employ the standard RMF Lagrangian with the exchange of σ, ω and ρ mesons between the nucleons. The corresponding Lagrangian density which describes the nucleons as Dirac spinors moving in meson fields is given by

$$\mathcal{L} = \bar{\psi}(\not{p} - g_\sigma \sigma - g_\omega \not{\omega} - g_\rho \vec{\rho}\vec{\tau} - \frac{1}{2}e(1-\tau_3)\not{A} - M)\psi$$
$$+ \frac{1}{2}\partial^\mu \sigma \partial_\mu \sigma - U(\sigma) + \frac{1}{2}m_\omega^2 \omega^\mu \omega_\mu + \frac{1}{2}m_\rho^2 \vec{\rho}^\mu \vec{\rho}_\mu \quad (2)$$
$$- \frac{1}{4}\Omega^{\mu\nu}\Omega_{\mu\nu} - \frac{1}{4}\vec{R}^{\mu\nu}R_{\mu\nu} - \frac{1}{4}F^{\mu\nu}F_{\mu\nu} + \frac{1}{4}g_4\left(\omega^\mu \omega_\mu\right)^2.$$

$U(\sigma)$ represents the conventional nonlinear σ potential. The vector self-coupling of ω-meson represented by the coupling constant $g_4$ was introduced in Ref. [22]. It was shown that the inclusion of the vector self-coupling of ω-meson provides improved shell effects in nuclei along the stability line [22]. This was the first attempt to calibrate the shell effects in nuclei in the RMF theory. Consequently, the force NL-SV1 [22] was introduced with a view to improve predictions of the ground-state properties of nuclei such as binding energies, charge radii and isotopes shifts of nuclei along the stability line and far away from it. Predictions made for stronger shell effects [23] near the *r*-process path using the vector self-coupling model NL-SV1 have since been confirmed experimentally [24-25]. Here, we have employed the same Lagrangian set NL-SV1 with the vector self-coupling of the ω-meson [22].

We have performed deformed RMF+BCS calculations using an expansion of fermionic and bosonic wave functions in 20 oscillator shells. The pairing is included using the constant gap approximation with pairing gap being taken from the widely used prescription of $\Delta_{n(p)} = 4.8 \, N^{-1/3}(Z^{-1/3})$ for open-shell nuclei [26]. The results of RMF+BCS calculations of charge radii obtained with NL-SV1 are shown in Fig. 1. The isotope shifts thus obtained show an increasing trend in going below N=50 except for N=44 where the experimental data is underestimated by the theory due to a relatively smaller deformation $\beta_2 \sim 0.10$ as compared to its neighbors. The experimental values are reproduced satisfactorily down to N=40 as well as above N=50. However, for the isotopes $^{74}$Kr (N=38) and $^{72}$Kr (N=36), there is a strong divergence of the theoretical values from the experimental data, though $\beta_2$ of these nuclei is significantly larger than that for $^{76}$Kr (N=40). The isotope shift for $^{70}$Kr (N=34) shows a further increase.

Theoretically, the rapid increase in the charge radius of $^{74}$Kr, $^{72}$Kr and $^{70}$Kr appears naturally as a consequence of the vicinity to the proton drip line. The BCS smearing of occupation probabilities across the Fermi surface which adjoins the continuum leads to a swelling of the charge radius. Evidently, a BCS description of nuclei close to the proton drip line runs contrary to the experimental data. The gulf between the experimental and theoretical values increases as one approaches the proton drip-line. It should be mentioned that the isotope shifts of Kr and Sr isotopes were investigated in a previous work [27] within the RMF+BCS formalism. In earlier times calculations were performed with 12 oscillator shells due to computing limitations. This gave rise to an apparently good agreement with the data [27]. Comparing this with the present realistic RMF calculations with 20 shells, one can see that this agreement is fortuitous. Clearly, 12 shells do not suffice to encompass the configuration space and hence underestimate the charge radii of neutron-deficient nuclei significantly. Due to this reason, the nuclear properties compiled in Ref. [28] using 12 oscillator shells are incorrect in the regions near the drip lines.

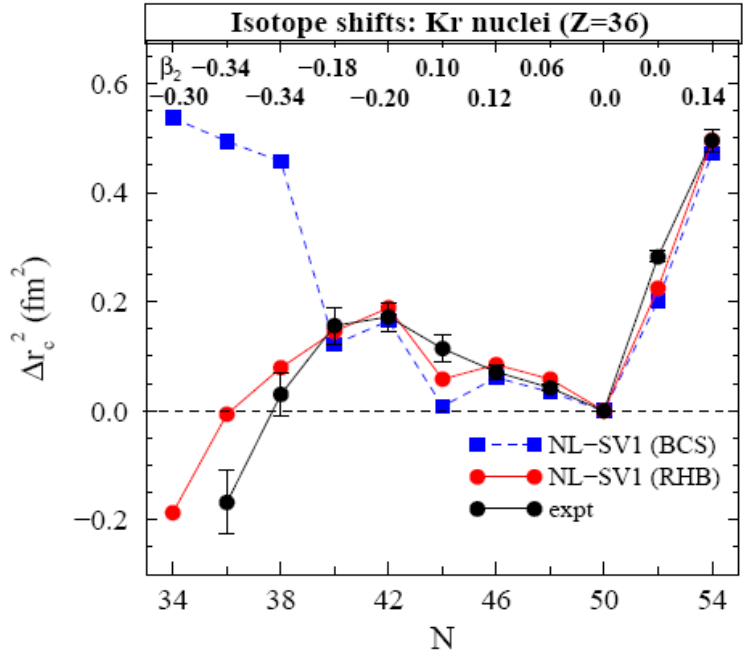

**FIGURE 1**: The isotope shifts $\Delta r_c^2$ of Kr chain obtained from deformed RMF+BCS and RHB calculations with NL-SV1. Deformation $\beta_2$ for the ground-state of nuclei obtained with NL-SV1 (RHB) are shown at the top of the figure. The experimental data [20] for $\Delta r_c^2$ are shown for comparison.

## B. The Deformed Relativistic Hartree-Bogoliubov theory

Notwithstanding the stark disagreement of the RMF+BCS calculations with the experimental isotope shifts, we have undertaken an inclusion of the pairing mechanism based upon Bogoliubov quasi-particle scheme in a deformed basis with a view to take into account the effects due to coupling of the Fermi surface to the continuum. The eigenstates are then eigenvectors of the generalized single-particle Hamiltonian containing the self-consistent mean-field and a pairing field $\Delta$ representing the particle-particle correlations. Staying within the Hartree approximation for the mean-field, the relativistic Hartree-Bogoliubov equations are written as [29]:

$$\begin{pmatrix} \hat{h}_D - m - \lambda & \hat{\Delta} \\ -\hat{\Delta}^* & -\hat{h}_D^* + m + \lambda \end{pmatrix} \begin{pmatrix} U \\ V \end{pmatrix}_k = E_k \begin{pmatrix} U \\ V \end{pmatrix}_k, \qquad (3)$$

where $\hat{h}_D$ is the single-particle Dirac Hamiltonian with quasi-particle energies $E_k$ and the chemical potential $\lambda$. Here, the pairing field

$$\Delta_{ab}(r,r') = \frac{1}{2}\sum_{cd} V_{abcd}(r,r') \kappa_{cd}(r,r') \qquad (4)$$

is the sum over matrix elements $V_{abcd}(\mathbf{r}, \mathbf{r'})$ of a two-body pairing interaction and the corresponding pairing tensor is defined by

$$\kappa_{cd}(r,r') = \sum_{E_k > 0} U_{ck}^*(r) V_{dk}(r') \ . \qquad (5)$$

The RHB equations are solved self-consistently in order to obtain eigensolutions and eigenvalues in a single quasi-particle basis. This is transformed to the canonical basis to obtain the desired observables. Here we use the Gogny force D1S [30] for the pairing channel. We have adjusted the strength of the pairing force in order to get a good agreement of the calculated binding energies with the experimental values for the chain of Sn isotopes. The highly superfluid Sn nuclei provide a best testing ground for calibration of the pairing strength. Consequently, the force NL-SV1 describes the ground-state binding energies over a large range of Sn isotopes very well.

We have performed deformed RHB calculations, with the pairing channel thus calibrated, for the Kr isotopes using an expansion of Dirac spinors and mesons fields into 20 shells of an axially deformed oscillator potential. In each case, minimizations have been achieved in the oblate and prolate regions of deformation. It is noticed that there is a shape-coexistence between an oblate and a prolate shape for several isotopes. The RHB calculations with NL-SV1 provide a good description of the ground-state binding energies of Kr isotopes with a few divergences within 0.20%.

The results obtained on the isotope shifts with NL-SV1 with the deformed RHB are shown in Fig. 1. The RHB results from N=48 towards N=40 show an improvement over the BCS values including that for N=44. The most interesting outcome of these calculations is the downward trend that arises for the isotopes below N=40, a picture that contrasts sharply with the BCS pairing. NL-SV1 overestimates the isotope shift (charge radius) of $^{72}$Kr (N=36) slightly. This is due perhaps to a slightly higher deformation ($\beta_2 = -0.34$) obtained theoretically than is the case. There is no datum available on $^{70}$Kr. We predict a negative isotope shift for this nucleus.

The case of the isotope $^{72}$Kr is noteworthy. It is an N=Z nucleus which participates significantly in the *rp*-process. Our results show that pairing correlations vanish both for protons and neutrons for this nucleus. This is due to Nilsson splitting of j-levels especially near the Fermi surface, which carves out a major shell gap both at N=36 and Z=36. Vanishing of the pairing implies that both N=36 and Z=36 become major shell closures in the deformed space. Thus, it renders the nucleus $^{72}$Kr as doubly magic. Accordingly, though these nuclei are close to the proton drip line, the complete absence of smearing engenders the decrease in the charge radius in stark contrast to the BCS result.

The isotopes $^{70}$Kr and $^{74}$Kr emerge with a major shell gap with vanishing pairing at the respective neutron numbers N=34 and N=36 in the ground state, thus exhibiting a neutron magicity only. This is, to a large extent, responsible for suppressing the charge radius of these isotopes.

## 3. DOUBLE MAGICITY IN THE N=Z RP-PROCESS NUCLEI

With the advent of the double magicity of $^{72}$Kr, we have explored the structure of the other N=Z *rp*-process nuclei. Results of the deformed RHB calculations with NL-SV1 show that not only $^{72}$Kr, but also the other N=Z nuclei $^{68}$Se, $^{76}$Sr and $^{80}$Zr exhibit larger shell gaps both at the respective neutron and proton numbers. These nuclei also display vanishing pairing correlations for neutrons and protons alike in the midst of a large deformation in the ground state. Strikingly, both protons and neutrons move in tandem coupled with the ensuing deformation in the ground-state thus creating large shell gaps in the deformed single-particle structure. This endows a double magicity to the nuclei $^{68}$Se, $^{72}$Kr, $^{76}$Sr and $^{80}$Zr with N=Z=34, 36, 38, and 40, respectively, in the deformed space.

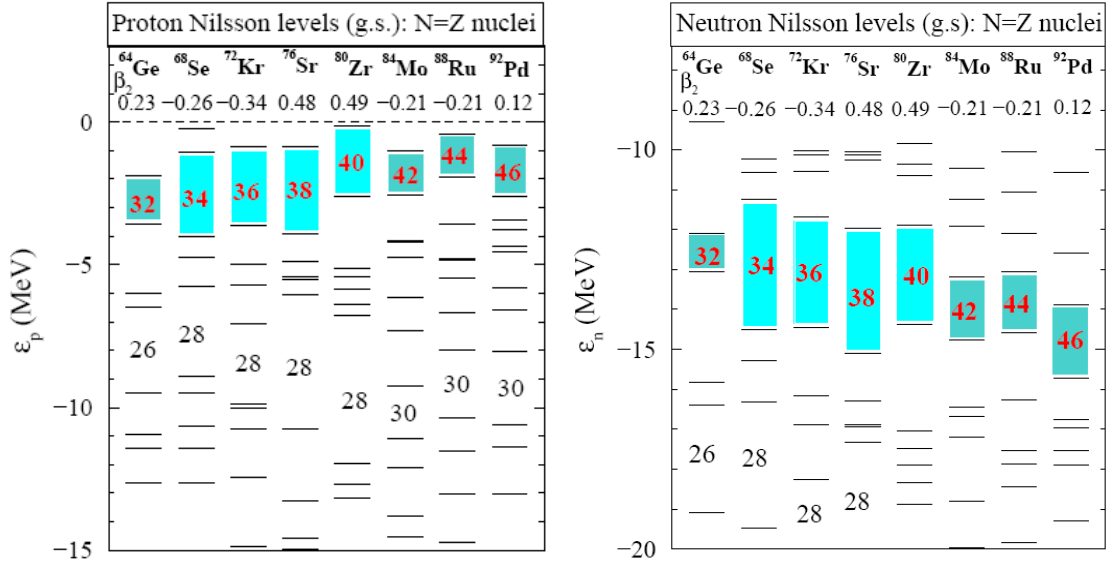

**FIGURE 2**: Proton and neutron single-particle levels for the ground-state of the N=Z nuclei obtained with deformed RHB theory using the force NL-SV1. Nuclei with the vanishing pairing correlations are highlighted with a larger shell gap in the columns 2-5.

The resulting single-particle structure of the N=Z nuclei is shown in Fig. 2, where the proton and neutron Nilsson single-particle levels obtained with the deformed RHB approach using NL-SV1 are displayed. The corresponding $\beta_2$ for the ground-state is indicated in the upper part of the figure. It is noteworthy that $^{76}$Sr and $^{80}$Zr exhibit a large prolate deformation in the ground state. This is consistent with the experimental values deduced [31,32]. The larger shell gaps at N and Z=34, 36, 38, and 40 can be seen conspicuously for the N=Z nuclei. These shell gaps correspond to a complete vanishing of the pairing correlations both for protons and neutrons. This constitutes the neutron and proton numbers N and Z=34, 36, 38 and 40 as shell closures in the deformed space. Consequently, nuclei with N=Z=34, 36, 38, and 40 are rendered as doubly magic.

In comparison, $^{64}$Ge (N=Z=32) shows shell gaps in neutrons and protons which do not give confidence for being a doubly magic nucleus. This is due to the non-vanishing pairing correlations and non-zero occupation numbers above the Fermi surface showing that $^{64}$Ge is not a magic nucleus according to our predictions. Doubts have been raised as to whether $^{64}$Ge is a waiting-point nucleus [33]. Other N=Z nuclei such as $^{84}$Mo, $^{88}$Ru, and $^{92}$Pd also show a well-deformed shape in the ground state and exhibit a shell gap both in protons and neutrons near the Fermi surface similar in magnitude to that in $^{64}$Ge with non-zero and yet significantly reduced pairing correlations and non-zero occupation numbers above the Fermi surface. These nuclei have been in contention for being waiting-points experimentally [3]. Our results, however, do not lend a magic or a doubly-magic character to these nuclei.

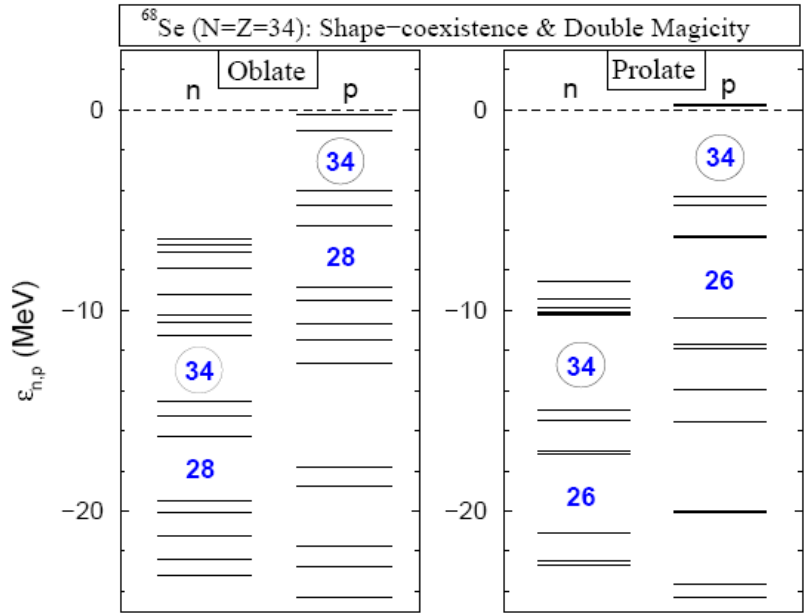

**FIGURE 3**. The neutron and proton Nilsson levels for the oblate (ground) and the prolate shapes coexisting in the nucleus $^{68}$Se (N=Z=34) obtained with NL-SV1 in the deformed RHB approach. The shell closures at N=Z=34 are shown by the encircled numbers.

## A. The Double magicity and Shape-coexistence in $^{68}$Se

The case of $^{68}$Se is peculiar. This nucleus has been shown experimentally to be a waiting-point nucleus beyond doubt [9]. Our results show that this nucleus has a second minimum with a prolate shape within 0.5 MeV of the oblate ground-state, thus exhibiting a shape-coexistence. Interestingly, the single-particle structure of the prolate state also exhibits a doubly magic character. This is illustrated in Fig. 3, where Nilsson levels for the oblate and the prolate state of the nucleus are shown. The single-particle structure of both the oblate (ground-state) as well as the prolate shape exhibits large shell gaps both at N=Z=34 (the shell closures being indicated by encircled numbers in Fig. 3) with the consequence that pairing correlations vanish completely both for neutrons and protons. This renders the nucleus $^{68}$Se as doubly magic both for the shape-coexisting oblate and the prolate states.

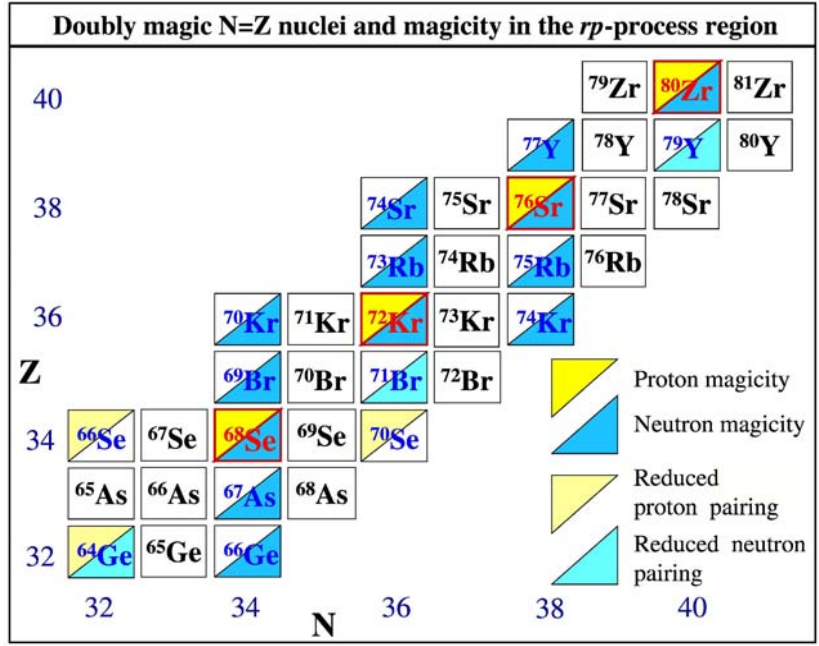

**FIGURE 4**: The N=Z nuclei exhibiting double magicity along the rp-process path. Nuclides exhibiting neutron magicity can be seen in the columns with N=34, 36, 38 and 40. A few cases with the reduced albeit non-zero pairing for neutrons and protons are also shown.

## B. The Island of Magicity in the RP-Process Region

We have investigated the single-particle structure of other nuclei in the vicinity of the *rp*-process path for N and Z=32-40. Deformed RHB calculations have been performed for sets of nuclei as shown in Fig. 3. As is prevalent in this region, ground state of nuclei leads to a significant deformation for most of the nuclei explored. Some cases also show a shape-coexistence between an oblate and a prolate shape. In Fig. 4 we show the nuclei which exhibit a neutron and/or proton magicity. The four N=Z nuclei $^{68}$Se, $^{72}$Kr, $^{76}$Sr, and $^{80}$Zr with double magicity stand out conspicuously.

We find that there is a preponderance of neutron magicity at several neutron numbers. This includes isotones with N=34, 36, 38 and to a limited extent with N=40. Surprisingly, a large number of nuclei in this region are amenable to a larger neutron shell gap at the Fermi surface with the vanishing pairing correlations. A few cases with significantly reduced neutron pairing correlations at N=32, 36 and 40 with the pairing energy of ~ −1 to −2 MeV are also shown in Fig. 4. In contrast, there are no cases of proton magicity found other than those of four N=Z doubly magic nuclei. Comparatively, the isotopes of $^{64}$Ge, $^{66}$Ge (Z=32) and $^{70}$Se (Z=34) exhibit proton shell gaps with non-zero yet significantly reduced proton pairing energy. Thus, proton magicity in this region is rather subdued as compared to neutrons.

## C. The Magicity and the Waiting-Point Nuclei

The *rp*-process region is encountered between the two magic numbers Z=28 and Z=50. Being the mid-shell region, it is prone to significant deformations of nuclei. Our revelation of the double magicity of $^{68}$Se, $^{72}$Kr, $^{76}$Sr, and $^{80}$Zr and especially the magicity of Z=34, 36, 38 and 40 in the deformed space comes out as a surprise. It may be noted that shell gaps in the single-particle spectrum of nuclei in the deformed space were predicted earlier on in the context of high-spin structure of nuclei [34]. However, a magicity of these numbers in the deformed space has eluded so far. Our analysis of the isotope shifts of Kr nuclei near the proton drip-line provides a first evidence of the magic character of the above nuclei.

It is interesting to note that peaks in abundances of masses at A= 64, 68, 72, 76, and 80 were obtained in *rp*-process calculations [4] in an x-ray burst, testifying to an important role played by the N=Z=32-40 nuclei. A double magicity of nuclei discerned in our work reinforces the waiting-point character of the *rp*-process nuclei similar to what is common for the waiting-point nuclei in the neutron-rich *r*-process region. It would be interesting to investigate as to how this attribute would influence the β-decay half-lives of nuclei and what would its effect on x-ray flux emanating from x-ray binaries be?

## 4. CONCLUSIONS

On the basis of our investigations of the experimental isotope shifts of Kr nuclei near the proton drip-line within the framework of the relativistic Hartree-Bogoliubov theory, we have found that N=Z *rp*-process waiting-point nuclei $^{68}$Se, $^{72}$Kr, $^{76}$Sr, and $^{80}$Zr exhibit double magicity in the deformed space.

Investigating the single-particle structure of the other nuclei in the *rp*-process region with N and Z=32-40, it is shown that the creation of large shell gaps in the deformed space in conjunction with the vanishing pairing correlations renders a preponderance of nuclides exhibiting a neutron magicity at N=34, 36, 38 and 40. This shows that the *rp*-process region close to the N=Z line is akin to an island of neutron magicity and double magicity. Comparatively, magicity of protons is subdued except that found for the four doubly-magic N=Z nuclei.

We have observed the unique case wherein the double magicity coexists with the shape-coexistence of oblate and prolate shapes in $^{68}$Se. This implies that the single-particle structure for $^{68}$Se rearranges itself so as to carve out a double magicity both for the oblate (ground) and the prolate states of the nucleus coexisting together.

## ACKNOWLEDGMENTS

This work is supported by the Project SP04/04 of the Research Administration, Kuwait University. MMS would like to thank Hendrik Schatz for useful discussions.